\documentclass[prl,aps,nobalancelastpage,amssymb,twocolumn,citeautoscript]{revtex4}

\usepackage{graphicx,multirow}

\begin{document}
\preprint{PRL/Xu \textit{et al}.}

\title{Evidence for two energy gaps and Fermi liquid behavior in SrPt$_2$As$_2$ superconductor}

\author{ Xiaofeng Xu$^1$\footnote[1]{Electronic address: xiaofeng.xu@hznu.edu.cn}, B. Chen$^1$, W. H. Jiao$^2$, Bin Chen$^3$, C. Q. Niu$^1$, Y. K. Li$^1$, J. H. Yang$^1$, A. F. Bangura$^4$,  Q. L. Ye$^1$, C. Cao$^1$,  J. H. Dai$^1$, Guanghan Cao$^2$\footnote[2]{Electronic address: ghcao@zju.edu.cn}, and N. E. Hussey$^5$}

\affiliation{$^{1}$Department of Physics, Hangzhou Normal University, Hangzhou 310036, China\\
$^{2}$State Key Lab of Silicon Materials and Department of Physics, Zhejiang University, Hangzhou 310027, China\\
$^{3}$Department of Physics, University of Shanghai for Science $\&$ Tehcnology , Shanghai, China\\
$^{4}$RIKEN(The Institute of Physical and Chemical Research), Wako, Saitama 351-0198, Japan.\\
$^{5}$H. H. Wills Physics Laboratory, University of Bristol, Tyndall Avenue, BS8 1TL, United
Kingdom\\}

\date{\today}

\begin{abstract}
We report a detailed calorimetric study on single crystals of the 5$d$-transition metal pnictide
SrPt2As2 with a superconducting critical temperature $T_c$ $\sim$5K. The peculiar field dependence
of the electronic specific heat coefficient $\gamma$ can be decomposed into two linear components.
Moreover, the temperature evolution of the electronic specific heat below $T_c$ is best described
by a two-gap model. These findings suggest that two energy gaps are associated with the
superconductivity. In parallel, we show that the spin-lattice relaxation time $T_1$, through
nuclear magnetic resonance measurement, obeys the so-called Korringa relation well. This, along
with the $T^2$ dependence of resistivity at low temperatures, points to a Fermi liquid ground state
in this material.
\end{abstract}

\maketitle

A central issue in the field of superconductivity is to elucidate the origin of the pairing
interaction, which in turn is intimately related to the pairing symmetry and the gap structure
$\Delta$(k). Notably, nodal $d$-wave superconductivity with $d_{x^2-y^2}$ pairing symmetry in
cuprates is generally believed to originate from the generic spin fluctuations in CuO$_2$
planes\cite{Lee06}. While in iron-based pnictides, the role played by antiferromagnetic spin
fluctuations is largely dependent on the strength of the iron 3$d$ electron correlation and remains
controversial \textit{albeit} a sign-reversing $s_\pm$ gap structure and multiple energy gaps have
been reported\cite{Hirschfeld11,Hu12}. In this regard, it is of fundamental importance to identify
the gap structure in understanding the underlying mechanism for the superconducting pairing glue.

Recently, motivated by the discovery of high $T_c$ superconductivity in ThCr$_2$Si$_2$-type
pnictides $A$Fe$_2$As$_2$ (where $A$ represents alkaline-earth metals)\cite{Rotter08}, a
$5d$-transition metal platinum-based 122 arsenide SrPt$_2$As$_2$ was found to be superconducting
below $T_c$$\sim$ 5K\cite{Imre07,Nohara10,Wang12}. In contrast to other 122 Fe-based
superconductors, this iron-free SrPt$_2$As$_2$ adopts a different CaBe$_2$Ge$_2$-type structure.
Its structure can be viewed as consisting of Pt$_2$As$_2$ tetrahedral layers alternating with
As$_2$Pt$_2$ layers stacked along the $c$-axis, the former layers with the Pt ion located in the
center of each As tetrahedron and the latter layer the opposite\cite{Shein11}. Remarkably, in
analogy to Fe-based pnictides, the SrPt$_2$As$_2$ compound also shows a structural phase transition
at $\sim$470K, which is associated with charge-density-wave (CDW) formation\cite{Wang12}. In spite
of these interesting discoveries, the nature of the low-lying quasiparticle excitations and the
pairing symmetry have yet to be addressed, in particular the role of the electron-phonon
interaction.

In this context, we investigate the superconductivity of single crystals SrPt$_2$As$_2$ via
detailed heat capacity measurement, a bulk probe of the low-lying quasiparticle excitations. Its
calorimetric responses, including the field evolution and the temperature dependence of the
quasiparticle specific heat, are overall consistent with a scenario of two $s$-wave superconducting
gaps opening on different sections of the Fermi surface. We therefore demonstrate that
SrPt$_2$As$_2$, along with the textbook example MgB$_2$\cite{Bouquet02}, is another prototypical
two-gap superconductor. In addition, we show from nuclear magnetic resonance measurement (NMR) that
spin fluctuations play only a minor role here. The observation of the Korringa law in this material
is consistent with $T^2$ resistivity at low temperatures, both pointing to a Fermi liquid ground
state.

Single crystals of SrPt$_2$As$_2$ was synthesized by self-melting technique, following the
procedure described in Ref. \cite{Wang12}. Good single crystallinity of the as-grown samples was
then confirmed by x-ray diffraction. For the transport measurements, the sample was cut into a
bar-like shape with the longest dimension along the basal plane. The specific heat measurement was
performed on a large piece of single crystal of weight 1.3 mg using a commercial Quantum Design
PPMS-9 system. The thermometer on the calorimeter puck was well calibrated prior to the
measurements in various magnetic fields used in this study between 10 K down to $^3$He temperature.
The addenda was determined in a separate run. We also performed $^{195}$Pt NMR study to investigate
the role of spin fluctuations and the Fermi liquid behavior in this material.

\begin{figure}
\includegraphics[width=9.5cm,keepaspectratio=true]{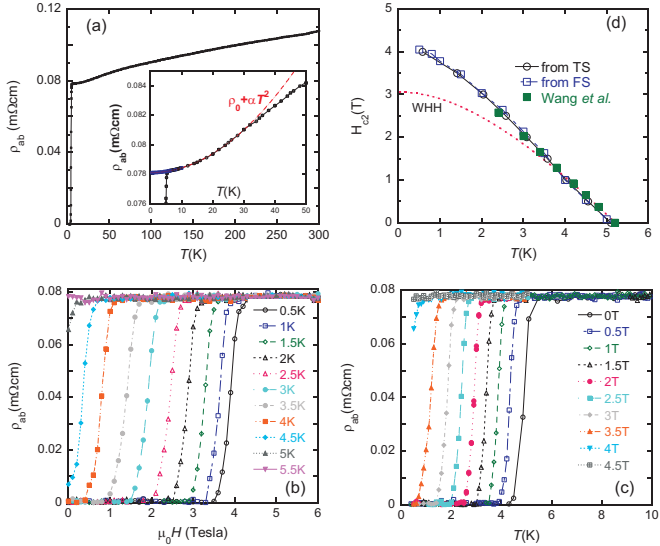}
\caption{(Color online) (a) The temperature dependence of the zero-field in-plane resistivity. The
inset zooms in the low temperature part with blue squares representing the data extracted from the
field sweeps at fixed temperatures (see text). The red line is the fit to $\rho_0+\alpha T^2$. (b)
and (c) show the field sweep (FS) and the temperature sweep (TS) respectively, $H\perp c$, with the
extracted $H_{c2}$ presented in (d), where the data from Fang \textit{et al.} are also plotted for
comparison\cite{Wang12}. The red dashed line in (d) is the WHH fit of $H_{c2}$\cite{WHH66}.}
\label{Fig1}
\end{figure}

Figure 1 encapsulates the zero-field resistivity and the extracted upper critical field $H_{c2}$ of
SrPt$_2$As$_2$. The temperature dependence of the resistivity is shown in part (a) of this figure
and is typical of the behavior reported in the literature, i.e. metallic behavior persisting to low
temperature where superconductivity eventually emerges below $T_c$$\sim$5.2K. The associated
$H_{c2}$ (for $H\perp c$) was then determined from both field and temperature sweeps, as shown in
Fig.1(b) and Fig.1(c), respectively. The 90$\%$ criterion was used to determine $H_{c2}$, i.e., the
field(temperature) at which $\rho$ reaches 90$\%$ of $\rho_n$, the resistivity of the normal state.
The as-drawn $H_{c2}$ was then depicted in Fig.1(d) in which the data from Fang \textit{et al}. was
also reproduced for comparison\cite{Wang12}. Clearly, our data overlap with those of Fang
\textit{et al.} down to the lowest temperature they studied, and persist in going up to 4 Tesla at
0.5K. We fit our data with conventional one-band Werthamer-Helfand-Hohenberg (WHH)
theory\cite{WHH66}, as displayed in Fig.1(d), by taking the initial slope near $T_c$, ($d
H_{c2}^{\perp c}/dT)$$\mid_{T=Tc}$=0.92T/K. However, this one-band WHH theory can not fit our low
temperature data satisfactorily. The discrepancy may be ascribed to many reasons, one of which is
that the observed superconductivity is not simply from one band but of multi-band
nature\cite{Hunte08}.
Here, we uncover another interesting feature of its normal state transport. As we see from
Fig.1(b), there is almost no magnetoresistance in the normal state of SrPt$_2$As$_2$, evidenced
from the flat feature of $\rho(H)$ curves above $H_{c2}$. Therefore, $\rho(9T)$ data, or
equivalently the intercept at $H=0$ by linearly extrapolating the $\rho_{ab} (T, H)$ above
$H_{c2}$\cite{Hussey09}, plotted as blue squares in the inset of Fig. 1(a), will represent the
genuine normal state resistivity at each temperature indicated. We find that the normal state
resistivity of SrPt$_2$As$_2$ is well fitted by $\rho_0+\alpha T^2$ up to $T \sim$ 32K, a hallmark
of the Fermi liquid ground state of a metal.

\begin{figure}
\includegraphics[width=9cm,keepaspectratio=true]{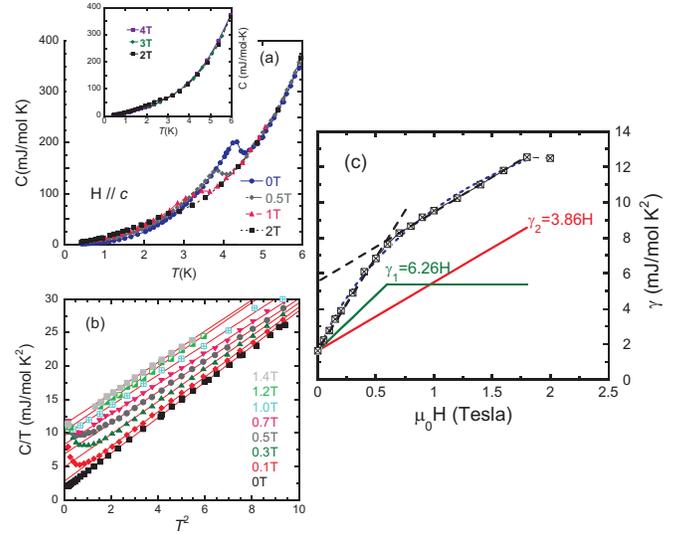}
\caption{(Color online) (a) The raw heat capacity data under various fields. The inset shows that
the data above 2 Tesla overlap with each other, indicative of the normal state of the sample under
these fields. (b) The plot of $C/T$ versus $T^2$ enables us to determine the $\gamma$ and $\beta$
as $C/T=\alpha+\beta T^2$. The red lines are the fits to this expression. The Debye temperature
estimated here is $\sim$160K, slightly lower than what Fang \textit{et al.} obtained\cite{Wang12}.
(c) shows the $\gamma$ coefficient as a function of field. This dependence was then decomposed into
two linear components $\gamma_1$ (Green) and $\gamma_2$ (Red) with the slope indicated in the
Figure. The blue dashed line is the fit to $H$log($c/H$), corresponding to $\gamma(H)$ for a dirty
$d$-wave superconductor.}
    \label{Fig2}
\end{figure}

The specific heat data at different fields are given in Fig.2(a). A clear heat capacity anomaly is
observed at $\sim$4.6 K in zero field. With increasing fields along the $c$-axis, this anomaly is
gradually suppressed to lower temperatures and ultimately disappears at $\sim$2 Tesla. As can be
seen in the inset, the data above 2 Tesla all collapse into a single curve. The plot of $C/T$
versus $T^2$, Fig.2(b), allows us to determine the Sommerfeld coefficient $\gamma$ and phonon
contribution $\beta T^3$. It is also noted that in the weak fields, there are upturns in $C/T$ at
very low temperatures, the origin of which is unknown to us. It may be attributed to the Schottky
specific heat due to the Zeeman splitting of the proton nucleus\cite{Nakazawa00}. The resultant
$\gamma$ as a function of field is summarized in Fig.2(c).

Fig. 2(c) displays one of the key findings of this paper, namely, the resultant $\gamma$ can be
broken down into two linear-in-field lines up to $H_{c2}$. As can be seen, $\gamma$ is linear in
fields up to $\sim$0.6T where it changes the slope but keeps growing linearly with fields and
finally saturates at $\sim$1.8 T. It is well known that in a dirty type-II $s$-wave superconductor,
$\gamma$ increases linearly with field due to the fact that the thermal excitations mainly arise
from the vortex core contribution in $s$-wave superconductor and the number of the vortices grows
linearly in $H$. In the clean limit of $d$-wave superconductors, however, $\gamma(H)$ scales as
$\sqrt H$ owing to the Doppler shift of the quasiparticle spectrum near the line nodes of the
gap\cite{Matsuda06}. Here in SrPt$_2$As$_2$, neither simple $s$-wave nor clean $d$-wave models
explain the observed $\gamma(H)$ curve. Fits to a model of dirty d-wave superconductivity, where
$\gamma$ scales as $H$log($c/H$) in weak fields \cite{Hussey02}, are shown in Fig. 2(c) and work
reasonably well. However, this is not compatible with other experimental data as shown below
\cite{footnote1}. Instead, reminiscent of other prototypical two-gap superconductors like
MgB$_2$\cite{Bouquet02} and Chevrel Phases\cite{Petrovic11}, where $\gamma(H)$ shares a wealth of
similarities with the present compound, the peculiar shape of $\gamma(H)$ here can also be
understood in a scheme of a two-band model. Consequently, in Fig. 2(c), we decomposed $\gamma(H)$
into two linear terms $\gamma_1$ and $\gamma_2$, with $\gamma_2$ linear up to 1.8T while $\gamma_1$
saturating at 0.6T. This corresponds to $\frac{\xi_1}{\xi_2}=\sqrt
\frac{{H_{c2}^{(2)}}}{{H_{c2}^{(1)}}}$=$\sqrt 3$. Assuming the same Fermi velocities on these two
bands, this immediately reveals $\Delta_2$ approximately 1.7 times of $\Delta_1$. It is also noted
that there is a finite $\gamma$ term $\sim$ 1.6 mJ/mol K$^2$ at zero field. We attributed this to
arising from the non-superconducting fraction of the sample, which corresponds to 13$\%$ of the
total normal carriers. By subtracting this contribution, we end up with the normal state $\gamma$
term for each band equal to $\gamma_n^{(1)}$=3.7 mJ/mol K$^2$ and $\gamma_n^{(2)}$=6.9 mJ/mol
K$^2$, respectively.

\begin{figure}
    \includegraphics[width=6cm,keepaspectratio=true]{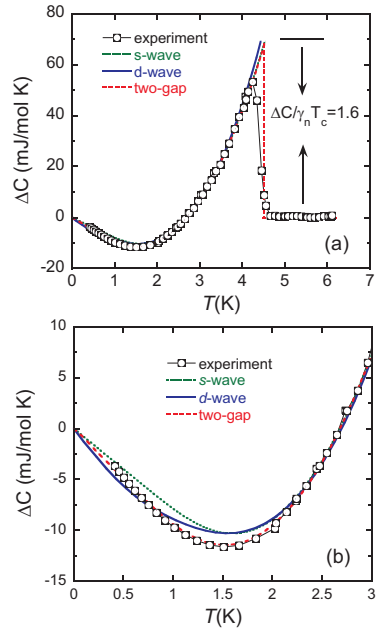}
\caption{(Color online) (a) The experimental data of $\Delta C= C$(0T)-$C$(4T) vs T, plotted with
the fits using three different gap functions. On this large $T$ scale, three different fits are
indistinguishable. However, with the expanded view at low temperatures, panel (b), it is evidently
seen the two-gap model fits the experimental data best. The resultant $\Delta C/\gamma_n T$ for
two-gap model, as indicated in (a), is equal to 1.6, close to the weak-couple BCS value of 1.43.}
    \label{Fig3}
\end{figure}

Considering a fraction of sample still being non-superconducting, we subtract the zero-field data
with those of 4T under which we believe the whole sample enters into the normal state, $\Delta
C=C$(0T)-$C$(4T). Since phone heat capacity is field independent, $\Delta C$ removes the phonon
contribution and the non-superconducting part, yielding $\Delta C=C_{es}-\gamma_sT$, where $C_{es}$
is the quasi-particle contribution and $\gamma_s$ here means the normal state Sommerfeld
coefficient of the superconducting part. As plotted in Fig. 3(a), the heat capacity jump at $T_c$
is best resolved in this fashion.

We analyzed our data by fitting $\Delta C$ with different gap functions. In BCS theory, the entropy
$S_{es}$ in the superconducting state is written as \cite{Owen07}

\begin{eqnarray}
S_{es}=-\frac{3\gamma_n}{k_B\pi^3}\int_0^{2\pi}\int_0^\infty[(1-f)ln(1-f)+flnf]d\varepsilon
d\phi,\label{eqn:one}
\end{eqnarray}

\noindent where $\gamma_n$ is the normal state $\gamma$, and $f$ stands for the quasi-particle
occupation number $f$=(1+$e^{E/k_B T}$)$^{-1}$ with $E$=$\sqrt {\varepsilon^2+\Delta^2(\phi)}$.
$\Delta(\phi)$ is the angle dependence of the gap function. For conventional $s$-wave, this is
angle-independent and for a standard $d$-wave, $\Delta(\phi)=\Delta(T)cos(2\phi)$. In addition, we
used the well-established $\alpha$ model to describe the temperature dependence of the gap
function, in which $\Delta(\phi,T)=\alpha\Delta_{\texttt{BCS}}(\phi,T)$\cite{Padamsee73}. Here
$\Delta_{\texttt{BCS}}(\phi,T)$ is the weak coupling BCS gap function. Therefore, the gap is
assumed to take the BCS-like form with the magnitude multiplied by a dimensionless parameter
$\alpha$, which gives the strength of the (electron-boson) coupling. The specific heat is
thereafter calculated by $C_{es}=T(\partial S/\partial T)$. For the two-gap fitting, two sets of
$\gamma_n$ and $\alpha$ are used for each gap respectively.

Given the small amount of sample being non-superconducting, we set $\gamma_n$ as an adjustable
parameter. Hence, the free parameters in the fitting are $\gamma_n$, $\alpha$ and $T_c$, similar to
the method used in Ref. \cite{Owen07}. At first sight, from Fig.3(a), all three gap functions fit
the experimental data equally well. However, the close-up view at the low temperature presented in
Fig. 3(b) immediately distinguish these three: The simple $s$-wave could not fit the data
satisfactorily nor the $d$-wave gap; Instead, the two $s$-wave gap functions best captures the
temperature evolution of $\Delta C$ in the whole temperature range below $T_c$. The corresponding
fitting parameters are listed in the Table I.

We now examine the two-gap fitting parameters to see its consistency with the picture we got thus
far. First of all, the $\gamma$ coefficients for each band obtained from the fitting are 3.44 and
6.32 mJ/mol K$^2$ respectively, in agreement with 3.7 and 6.9mJ/mol K$^2$ extracted from Fig. 2(c).
In addition, the ratio of $\alpha$ (or equivalently $\Delta$) values between these two gaps is
1.52, close to 1.7 obtained above by assuming the same Fermi velocity for each band. The small
discrepancy may duly arise from the difference in the Fermi velocities of these two bands. Finally,
this two-gap superconductivity is not conflicting with the band structure calculation where the
Fermi surfaces consist of electron and hole pockets\cite{Shein11}.

\begin{table}
\centering \caption{The derived fitting parameters using Eqn. \ref{eqn:one} for three different gap
functions. $\gamma_n$ is in the units of mJ/mol K$^2$. }
\begin{tabular}{|c||c|c|c|c|c|}
\hline
$$ & $T_{C} $[K]$$ &\multicolumn{2}{|c|}{$\alpha$}&\multicolumn{2}{|c|}{$\gamma_n$}\\
 \hline
$s$-wave & 4.6 &\multicolumn{2}{|c|}{1.2098}&\multicolumn{2}{|c|}{7.97}\\
\hline
$d$-wave &4.8&\multicolumn{2}{|c|}{1.3088}&\multicolumn{2}{|c|}{12.23}\\
 \hline
\multirow{2}{*}{two-gap}&\multirow{2}{*}{4.6}& $\alpha_{1}$ & $\alpha_{2}$ & $\gamma_n^{(1)}$ &$\gamma_n^{(2)}$\\
\cline{3-6}
 & &0.83508&1.2719 & 3.44 &6.35\\
\hline
\end{tabular}
\end{table}

Fig. 4(a) shows the Knight shift of $^{195}$Pt below 50 K, obtained from the peak resonance field
as exemplified at 30 K in the inset. The $T$-independence of the observed Knight shift indicates
the nearly constant spin susceptibility, consistent with the bulk susceptibility measurement which
is predominated by $T$-independent Pauli paramagnetism (data not shown here). The spin-lattice
relaxation rate 1/$T_1$ was measured at the peak position of the field-swept NMR spectrum, and
determined by fitting the recovery curve to a single exponential function ($\varpropto$
exp($-\frac{t}{T_1})$) at each fixed temperature. As plotted in Fig. 4(b), 1/$T_1$ decreases
monotonically with temperature in the whole temperature range studied. The linear temperature
dependence of 1/$T_1$, known as the Korringa relation $T_1T$=const, is generally believed to arise
from the Fermi liquid excitations in a metal. This linear relation yields $T_1T$ = 0.23 s K for
$^{195}$Pt. Interestingly, this type of behavior had also been observed in MgB$_2$ by $^{25}$Mg and
$^{11}$B NMR measurements\cite{MgNMR,BNMR}. The validity of the Korringa relation in this compound,
together with its $T^2$ resistivity at low temperatures, is indicative of a Fermi liquid ground
state.

\begin{figure}
    \includegraphics[width=7.0cm,keepaspectratio=true]{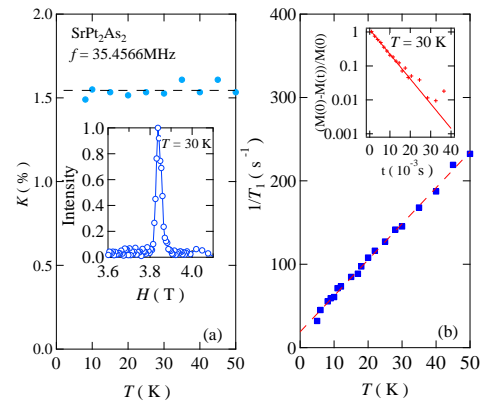}
\caption{(Color online) (a) Temperature dependence of $^{195}$Pt Knight shift for SrPt$_2$As$_2$.
As an example, the inset shows the field-swept $^{195}$Pt-NMR spectra at 30 K measured with the
fixed frequency of 35.4566 MHz. (b) Temperature dependence of the $^{195}$Pt nuclear spin-lattice
relaxation rate 1/$T_1$. The dotted line is the fit to the Korringa relation $T_1T$ = const. The
inset shows the typical recovery curve measured at 30 K. } \label{Fig4}
\end{figure}

In summary, based on the experimental probes used in this study, namely, the (magneto)transport,
the specific heat as well as the NMR measurements, a coherent picture can be achieved within the
framework of two-gap, $s$-wave, BCS-like superconductivity developed on the background of a Fermi
liquid. The observed superconductivity in this material is seemingly difficult to be associated
with any spin fluctuations, while the role of the orbital fluctuations is not clear here.
Theoretically, the frustration between the electron-hole interband interaction and the
electron-electron intraband scattering can induce a nodal gap\cite{Kasahara12,Kuroki09}. However,
this does not seem to be seen here. The observed two-gap energy structure, like the case of
MgB$_2$, seems to suggest the coupling between different bands is rather weak\cite{Bouquet02}.
Significantly, it is well known that in 5$d$ electron systems, the Coulomb interaction is
relatively weak while the spin-orbit coupling could play an essential role in the electronic
structure and their possible superconductivity. Yet, this role may not be always as prominent as
previously thought, because it can be weakened near the Fermi surface as evidenced in another
platinum-based pnictide SrPt$_3$P \cite{Cao12}, where the superconductivity with $T_c$$\sim$8.4K
and strong-coupling to bosonic modes were observed \cite{Takayama12}. Finally, the resultant
Kadowadi-Woods ratio (KWR) from our measurements is found to be
$\sim$20$\mu\Omega\cdot$mol$^2$$\cdot$K$^2$/J$^2$. This is two orders of magnitude greater than the
value typically found for transition metals, and close to those for heavy fermions and strongly
correlated oxides. Taken at face value, this potentially indicates strong electron-electron
correlations. However, Hussey \cite{Hussy05} and Jacko \textit{et al}. \cite{Jacko09} both point
out that the absolute value of KWR does not necessarily reveal anything about the strength of
electronic correlations since material specific parameters (e.g., dimensionality, carrier density,
multi-band effects) may be responsible for an enhanced KWR.

The authors would like to acknowledge valuable discussions with Xiaofeng Jin, Y. Matsuda, T.
Shibauchi and C. M. J.  Andrew, and the technical support from X. X. Yang, H. D. Wang.
This work was supported by the National Natural Science Foundation of China. 

\bibliography{SrPt2As2}


\end{document}